\documentclass[10pt,conference]{IEEEtran}
\usepackage[utf8]{inputenc}
\usepackage[T1]{fontenc}
\usepackage{amsmath,amssymb}
\usepackage{algorithm}
\usepackage{algorithmic}
\usepackage{booktabs}
\usepackage{multirow}
\usepackage{xcolor}
\usepackage{graphicx}
\usepackage{tikz}
\usetikzlibrary{arrows.meta,positioning,fit,backgrounds,calc}
\usepackage{pgfplots}
\pgfplotsset{compat=1.18}
\usepgfplotslibrary{groupplots}
\usepackage{textcomp}
\usepackage{comment}
\usepackage[hidelinks]{hyperref}

\makeatletter
\DeclareRobustCommand{\IEEEauthorrefmark}[1]{%
  \smash{\textsuperscript{\footnotesize #1}}%
}
\makeatother

\begin{document}

\title{EviRCA: Decoupling Evidence Extraction from Reasoning for Microservice Root-Cause Analysis%
    \thanks{Title and author list are provisional.}}

\author{
    \IEEEauthorblockN{
        Yuhao Wang\IEEEauthorrefmark{1},
        Zhen Qin\IEEEauthorrefmark{2}\textsuperscript{*},
        Xingliang Wang\IEEEauthorrefmark{1},
        Guochang Li\IEEEauthorrefmark{1},
        Weize Li\IEEEauthorrefmark{3},
        and Shuiguang Deng\IEEEauthorrefmark{1}\textsuperscript{*}
    }
    \IEEEauthorblockA{
        \IEEEauthorrefmark{1}\textit{College of Computer Science and Technology, Zhejiang University}, Hangzhou, China \\
        \IEEEauthorrefmark{2}\textit{School of Software Technology, Zhejiang University}, Ningbo, China \\
        \IEEEauthorrefmark{3}\textit{China Telecom Cloud Computing Corporation}, Beijing, China \\
        \texttt{22421122@zju.edu.cn, zhenqin@zju.edu.cn, wangxingliang@zju.edu.cn, gcli@zju.edu.cn,} \\
        \texttt{liweize@chinatelecom.cn, dengsg@zju.edu.cn} \\
        \textsuperscript{*}Corresponding authors: Zhen Qin and Shuiguang Deng.
    }
}

\maketitle

\begin{abstract}
Root-cause analysis (RCA) is a critical yet labor-intensive task for maintaining modern microservice systems, making it an attractive target for applying large language models (LLMs)
Recent agentic approaches allow an LLM to iteratively explore raw telemetry by generating and executing code. However, this design asks a single model to simultaneously retrieve evidence, localize faults, and infer root causes over large volumes of heterogeneous telemetry, leading to high computational cost, unstable behavior, and limited diagnostic accuracy.
However, the telemetry data consists of numeric metrics, structured traces, and machine-generated logs, which are not appropriate for being directly processed by LLMs. Therefore, this work presents EviRCA, a framework for LLM-based RCA that decouples deterministic evidence extraction from LLM reasoning.
A system-agnostic extraction stage converts raw metrics, traces, and logs into a compact set of faithful multimodal evidence cards, while the LLM reasons only over these structured observations through a small set of predefined read-only tools, without accessing raw telemetry or executing code.
We evaluate EviRCA on OpenRCA, a benchmark built from real, heterogeneous telemetry across three enterprise systems.
EviRCA achieves a Correct rate of 40.6--43.9\% across two different LLMs, substantially outperforming prior OpenRCA baselines that achieve up to 15.2\%, while reducing token consumption by 15--26$\times$ and execution time by 3--20$\times$.
Moreover, EviRCA successfully solves hard cases requiring simultaneous processing of time, components and reasons, a setting in which previous approaches reported near-zero performance.
Our process-level failure analysis further shows that the binding difficulty lies in judging evidence that the extraction stage has already surfaced, instead of searching for it.
This suggests that the effectiveness of LLM-based RCA highly depends on the quality of evidence extraction.
\end{abstract}


\section{Introduction}
\label{sec:intro}

Cloud-native microservice architectures now underpin much of the software that businesses and users depend on~\cite{Dragoni2017}.
However, their scale and constant change make failures frequent and costly, demanding continuous on-call attention while risking substantial financial and reputational damage.
When a service degrades, on-call engineers must perform root-cause analysis (RCA) to trace user-visible symptoms back to the faulty component.
This process requires inspecting massive, heterogeneous telemetry data, including metrics, logs, and traces.
While traditional RCA algorithms, such as call-graph random walks, causal-graph construction and log-trace analysis, can be accurate, they heavily rely on pre-specified topologies, curated training data, or system-specific modeling, limiting their transferability to new deployments~\cite{10.1145/3501297, 10.1145/3736755}.

The emergence of large language models (LLMs) has fundamentally reshaped many software engineering tasks.
In domains such as code generation, issue resolution, and testing, agentic systems and benchmarks such as SWE-bench~\cite{jimenez2024swebench}, SWE-agent~\cite{3737916.3739517}, OpenHands~\cite{wang2025openhands}, and MetaGPT~\cite{hong2024metagpt} have driven rapid progress, and tools such as Copilot~\cite{copilot2026}, Codex~\cite{codex2026}, and Claude Code~\cite{claudecode2026} have integrated into daily development.
These successes share a common basis: their inputs are natural-language-friendly, such as source code and issue descriptions, and their outputs admit cheap, executable verification through tests.
This capacity for natural-language interaction also makes LLMs highly attractive for RCA, allowing engineers to state diagnostic goals intuitively without worrying about environment-specific algorithms.
Early LLM-based RCA approaches primarily treated the model as a text summarizer or classifier over incident reports, focusing on predicting broad root-cause categories or suggesting mitigations~\cite{10172904, 10.1145/3627703.3629553, 10.1145/3597503.3639081}.
These outputs, however, remain at the level of textual summarization or coarse category prediction, without localizing the fault in raw, live telemetry.
OpenRCA~\cite{xu2025openrca} provides a concrete formulation of the problem by introducing a goal-driven, reproducible benchmark built on heterogeneous telemetry collected from three enterprise systems, in which a method must recover a queried subset of a fault's time, component, and reason.
Its reference method, RCA-Agent, gives the LLM agency over the telemetry, letting it repeatedly write and run Python over the raw data in the ReAct style~\cite{yao2023react} until it commits an answer.
Yet, despite this operational freedom, the strongest LLM evaluated in the original study solves a mere $11.3\%$ of cases~\cite{xu2025openrca}.

This limited performance reflects a fundamental mismatch between LLMs and the nature of telemetry data.
Raw telemetry consists of high-volume, low-level signals including numerical metrics, structured traces, and machine-generated logs, which provide little semantic abstraction for language models.
Unlike natural language or source code, telemetry cannot be directly processed through pretrained linguistic representations, nor can conclusions be validated using inexpensive executable tests.
Consequently, requiring a single agent to detect anomalies, localize, and attribute over this raw telemetry within one loop is prohibitively expensive and brittle.
Such an unconstrained loop can easily consume nearly a million tokens per incident, which may produce hallucinated data points and schema fields while causing latency to grow with every step.
Furthermore, the traditional end-to-end evaluation protocol, which counts a prediction as correct only when every queried field is recovered exactly, obscures the underlying source of failure. 
A single pass/fail metric cannot distinguish failures caused by inadequate evidence retrieval from those caused by incorrect reasoning over evidence that has already been obtained.
Concurrent studies corroborate this conclusion: agentic RCA failures are largely shared across model capability tiers and rooted in the agent architecture itself, and end-to-end scoring conflates root-cause reasoning with unrelated skills such as anomaly detection and code generation~\cite{kim2026aiagentssystematicallyfail, Riddell2026StalledBA}.

Goal-driven RCA naturally decomposes into three subtasks: 
1) anomaly detection over heterogeneous telemetry, 2) candidate localization, and 3) semantic discrimination of the true root cause and its underlying failure reason.
In fact, the first two are deterministic signal-processing problems that can be solved reliably and reproducibly without an LLM, whereas only the final stage requires semantic reasoning.
We therefore argue that \emph{goal-driven RCA should focus not on discovering evidence through unconstrained exploration, but on accurately interpreting evidence once it has been retrieved}.
This motivates a more effective division of labor: deterministic components first compress raw telemetry into compact, faithful evidence, allowing the LLM to focus exclusively on high-level diagnostic reasoning.

Guided by this insight, we present EviRCA, a framework that explicitly separates these two responsibilities. 
A deterministic, system-agnostic stage first extracts compact yet faithful multimodal evidence from raw telemetry to formulate a set of evidence cards, while an LLM-based stage diagnoses the root cause solely based on this structured evidence through a restricted, read-only interface, without reading raw telemetry or executing code. 
In EviRCA, all system-specific knowledge is encapsulated in a lightweight declarative adapter, enabling the same reasoning engine to operate across heterogeneous systems without architectural changes.

This work makes the following contributions:
\begin{itemize}
    \item \textbf{We identify the principal bottleneck of goal-driven LLM-based RCA}. 
    Through process-level failure analysis and controlled ablations, we show that the dominant challenge is reasoning over surfaced evidence rather than searching raw telemetry with LLMs. 
    This provides a new understanding of how LLMs should be incorporated into telemetry-grounded diagnosis.
    \item \textbf{We propose EviRCA, a two-stage framework for goal-driven RCA}. It separates deterministic evidence extraction from LLM-based semantic reasoning, represents heterogeneous telemetry as compact multimodal evidence cards, and isolates system-specific knowledge through a declarative adapter so that the same engine generalizes across multiple enterprise systems.
    \item \textbf{We perform extensive experiments on the OpenRCA benchmark}. Across three enterprise systems and two LLMs from different model families, EviRCA improves the Correct rate to $40.6$--$43.9\%$ compared with at most $15.2\%$ for the baselines with matched models, while reducing token consumption by 15--26$\times$. Specifically, it successfully solves hard-tier cases requiring recovery of the complete tuple including time, component and reason, which is a setting in which previous methods report 0\% accuracy.
\end{itemize}

\section{Background and Motivation}
\label{sec:background}

\subsection{Goal-Driven Root-Cause Analysis}
\label{sec:background:task}

OpenRCA frames RCA as a goal-driven task over telemetry~\cite{xu2025openrca}.
Each case consists of a 30-minute window of telemetry and a natural-language query describing a failure that occurred within it.
The telemetry is multimodal.
Metrics are per-component time series of resource usage (CPU, memory, disk, network) together with application-level indicators such as request rate and latency.
Traces record the call graph between components, along with each call’s latency and, where recorded, its success or failure.
Logs, when available, are textual event streams that can surface failures metrics miss, such as a burst of garbage-collection messages.
A method must therefore fuse evidence across these modalities, because a fault invisible in one is often conspicuous in another.

The query asks for a subset of the root cause’s three attributes (time, component, and reason); a window may contain more than one fault.
Scoring is exact and all-or-nothing: components and reasons must belong to a closed, system-specific candidate set, and a case is judged Correct only when every queried attribute matches the ground truth (the precise metric is defined in Section~\ref{sec:eval:setup}).
This contract rewards a single precise decision over broad exploration, which is what makes the task challenging.

\subsection{Structural Heterogeneity Across Systems}
\label{sec:background:hetero}

\begin{table*}[t]
\centering
\caption{The three OpenRCA systems differ on every structural axis, so a method must absorb these differences.}
\label{tab:heterogeneity}
{\renewcommand{\arraystretch}{1.1}
\begin{tabular}{@{}l p{0.205\textwidth} p{0.205\textwidth} p{0.205\textwidth}@{}}
\toprule
 & \textbf{Bank} & \textbf{Telecom} & \textbf{Market} \\
\midrule
Cases                & 136 & 51 & 148 \\
Candidate levels     & pod & node, pod, service & node, pod, service \\
Candidate components & 14 pod (6 service) & 22 node, 8 pod, 13 service & 6 node, 40 pod, 10 service \\
Vertical topology    & none & node--pod co-location & encoded in pod names; failover \\
Modalities           & metric, trace, log & metric, trace (no log) & metric, trace, log\\
Golden signal        & synthetic service tests & single gateway (OSB) & per service \\
Reason types         & 8 & 5 & 15 \\
Time unit            & metric\,s, trace\,ms, log\,s & all ms & metric\,s, trace\,ms, log\,s \\
\bottomrule
\end{tabular}
}
\end{table*}

The three systems share the task but differ in nearly every structural respect (Table~\ref{tab:heterogeneity}).
First, they differ in their component hierarchy: Bank exposes a flat set of pods, whereas Telecom and Market are organized into node, pod, and service levels.
Consequently, the level at which a fault must be reported and the manner in which a symptom propagates between levels vary from system to system.
Second, they differ in the modalities that are actually present: Telecom lacks logs, and only Market exposes a service mesh, so the evidence available for fault attribution is not uniform.
Third, they differ in time encoding; timestamps are recorded in units that range from seconds to milliseconds across files, making raw comparison impossible without normalization.
Finally, they differ in how faults manifest and in how many possible reasons they admit.
For example, a failure may surface as a database on/off flag on Telecom, as garbage-collection logs on Bank, or as mesh-edge status codes on Market.
The number of distinct reasons also varies: 5, 8, and 15 for the three systems, respectively.
These differences are structural, not cosmetic.
A method that hard-codes the assumptions of one system will not transfer; therefore, handling all three systems simultaneously is the generalization challenge posed by this benchmark.

\subsection{Agentic RCA and Its Costs}
\label{sec:background:agentic}

OpenRCA includes a reference method, RCA-Agent, that gives the LLM broad latitude to explore the telemetry: a planner repeatedly instructs an executor, which writes and runs Python over the raw data in a ReAct loop~\cite{yao2023react}, inspecting and aggregating telemetry until it produces an answer.
This reflects a broader class of autonomous RCA agents that operate on raw or lightly structured telemetry through tools or generated code~\cite{10.1145/3627673.3680016, 10.1145/3663529.3663841, 10.1145/3701716.3715225, 11229957}.
Such open-ended exploration is flexible, but it requires the model to process large volumes of raw telemetry across many steps per case, making it costly and unstable---as we quantify in Section~\ref{sec:eval:rq2} and dissect in Section~\ref{sec:eval:rq4}.
EviRCA departs from this paradigm.
Instead of letting the model explore raw telemetry freely, it enforces a strict separation: a deterministic stage handles the detection and extraction that can be done reliably without an LLM, while the LLM is reserved for the final reasoning step that truly requires it.

\section{Approach Design}
\label{sec:design}

\subsection{Overview}
\label{sec:design:overview}
\begin{figure*}[t]
\centering
\IfFileExists{fig_overview_pdf.pdf}{%
  \includegraphics[width=\textwidth]{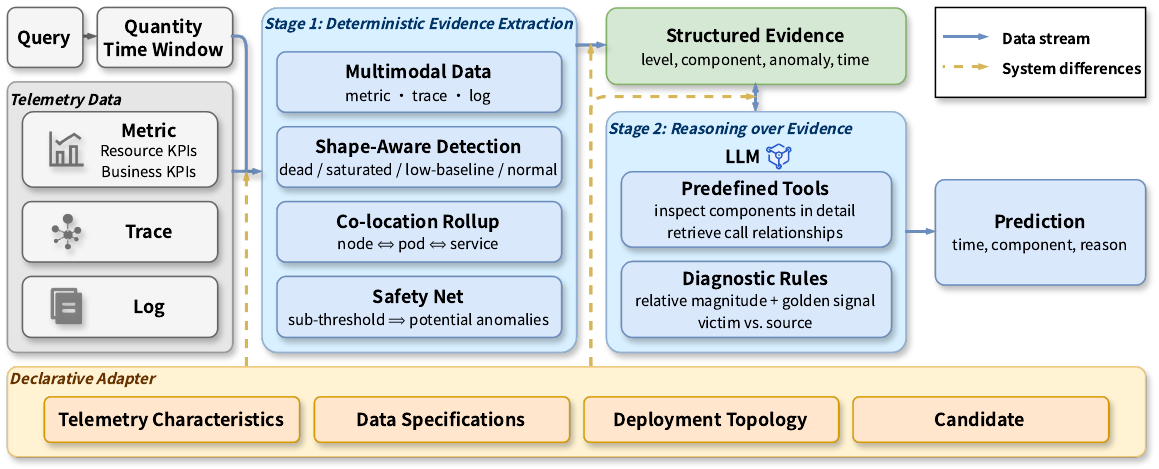}%
}{%
  \fbox{\parbox[c][3.2cm][c]{0.98\textwidth}{\centering\itshape
  [ fig\_overview.pdf not found --- place the exported overview figure beside main.tex. ]}}%
}
\caption{Overview of EviRCA.
Stage 1 reduces raw multi-modal telemetry to structured evidence cards through deterministic signal processing, 
and Stage 2 lets an LLM reason over them to predict the $\langle$time, component, reason$\rangle$ of each fault.
A Declarative Adapter supplies per-system specifications to both stages, keeping the engine system-agnostic.}
\label{fig:overview}
\end{figure*}

Goal-driven root-cause analysis over a microservice system decomposes into three sequential subtasks: 1) detecting anomalies in multi-modal telemetry, 2) forming candidate localizations, and 3) discriminating the true root cause and its reason from competing candidates.
Prior studies, such as RCA-Agent, collapse all three subtasks into one agentic loop: an LLM repeatedly writes and runs Python over the raw telemetry until it commits an answer.
EviRCA instead matches each subtask to the mechanism suited to it (Fig.~\ref{fig:overview}).
Because detection and evidence extraction are deterministic, repeatable signal-processing tasks, EviRCA assigns them to a system-agnostic stage that reduces raw telemetry to faithful, compact evidence.
Discrimination and attribution require semantic judgment over correlated candidates, so EviRCA assigns them to an LLM stage that reasons over the extracted evidence through a small set of predefined, read-only tools.

As shown in Fig.~\ref{fig:overview}, EviRCA takes two inputs: the multimodal telemetry recorded within the query window, and the natural-language query that names the target attributes to recover.
It then proceeds through two stages.
The first stage, Deterministic Evidence Extraction (Sec.~\ref{sec:design:evidence}), runs entirely in code and reduces the raw telemetry to a compact set of structured \emph{evidence cards}, each describing one anomalous (component, resource-type) pair.
The second stage, Reasoning over Structured Evidence (Sec.~\ref{sec:design:reasoning}), hands an LLM these cards and lets it discriminate the root cause through read-only tools, without ever reading raw telemetry or executing code.
Binding the two stages together is a Declarative Adapter (Sec.~\ref{sec:design:adapter}), a per-system specification of objective structure such as the telemetry layout, the candidate hierarchy, and the deployment topology, which both stages consult so that the engine itself contains no per-system logic.
Finally, EviRCA outputs a $\langle$time, component, reason$\rangle$ prediction for each fault, reconciled to the number of injected faults the case contains.

The following subsections describe the two stages and the adapter in turn.

\subsection{Deterministic Evidence Extraction}
\label{sec:design:evidence}

The purpose of the first stage is to provide the reasoning stage with something small, faithful, and complete to work with, in place of raw telemetry.
Its role in the pipeline is to perform the two subtasks that require no language model, anomaly detection and candidate localization, and to do so deterministically.
Doing this work in code rather than in the LLM brings three benefits that an agentic loop lacks.
The output is reproducible, since the same window always yields the same cards.
The model can no longer hallucinate data points or schema fields, since it never sees the raw series.
Finally, the evidence it passes downstream is already compact, which keeps the reasoning stage cheap.

The stage takes the raw telemetry in the query window, together with the structure declared by the adapter, and produces a set of evidence cards.
Each card records one anomalous (component, resource-type) pair: its onset time, the KPIs that support it, and a set of \emph{coarse} reason candidates.
Before any window is cut, EviRCA computes a robust statistical baseline for every component KPI over the full day, so that a short fault inside the window is judged against the component's own normal behavior.
The stage then runs four steps: shape-aware detection finds the anomalous series, multimodal card construction normalizes anomalies from every modality into the common card format, co-location rollup re-levels evidence to the granularity at which a fault must be reported, and a sub-threshold fallback guarantees that the output is never empty.
We describe each step in turn.

\textbf{Shape-aware detection.}
This step takes each component KPI's full-day series and its baseline and decides whether, and when, the series becomes anomalous inside the window.
A single anomaly test cannot serve every KPI, because operational telemetry contains series of very different shapes, and a test tuned for one shape misfires on another.
EviRCA therefore first classifies each series by the shape of its daily distribution and then applies a detector matched to that shape.
We distinguish four shapes, each with its characteristic signature and a correspondingly specialized detector:
\begin{itemize}
  \item \emph{Dead} series have effectively no variance and carry no signal, so they are discarded.
  \item \emph{Saturated} series sit persistently at the top of their range, with even their low percentiles close to the maximum, so an absolute upper threshold can never fire; such series are flagged and routed to the fallback, since their fault is invisible to a metric threshold and must be recovered through other evidence.
  \item \emph{Low-baseline} series rest near zero and move only in sparse spikes, so a purely relative test reads ordinary noise as a large excursion; they must therefore clear an absolute-magnitude floor on top of the relative criterion, which keeps a jump from zero to a negligible value from being reported as a fault.
  \item \emph{Normal} series are well behaved and admit a standard test for a sustained, significant excursion beyond a robust band around the baseline.
\end{itemize}
Across all shapes, an anomaly is accepted only when the excursion persists for several consecutive samples and is large relative to the band, which suppresses isolated spikes and marginal breaches.
Treating all four shapes alike would miss faults on the dead and saturated series and fabricate them on the low-baseline ones.

\textbf{Multimodal card construction.}
This step takes the anomalies surfaced from every available modality and normalizes them into one card format.
EviRCA runs a detector per modality: the shape-aware test above over metrics, a trace detector that flags failed or abnormally slow calls, and a log detector that flags telltale bursts such as a surge of garbage-collection messages signalling memory pressure.
Each detected anomaly becomes a card for its (component, resource-type) pair, annotated with coarse reason candidates drawn from the adapter's resource-type-to-reason map.
When a resource type is consistent with more than one reason, the card carries several candidates, and the choice among them is deferred to the reasoning stage.
Running every modality matters because a fault that is silent in one is often loud in another: a memory exhaustion that never trips a metric threshold still appears as a garbage-collection burst in the logs, and a node fault with no node-level metric still appears as latency in the traces.
Extracting all modalities is what makes the card set high-recall, so that the reasoning stage rarely has to search for evidence the extraction stage failed to surface.

\begin{table*}[t]
\centering
\footnotesize
\renewcommand\arraystretch{1.1}
\setlength{\tabcolsep}{2pt}
\caption{Tools available to the LLM reasoning stage.}
\label{tab:tools}
\begin{tabular}{@{}l p{0.14\textwidth} p{0.38\textwidth} p{0.28\textwidth}@{}}
\toprule
Tool & Input & Output & Description \\
\midrule
\texttt{list\_entries} & --- & Detected entries (component, type, onset, KPIs); weak leads if none & Starting leads from the deterministic layer \\
\addlinespace[1pt]
\texttt{get\_component} & component & Its entries, blind/saturated types, topology, available KPIs & Full evidence for one component \\
\addlinespace[1pt]
\texttt{get\_kpi\_series} & component, KPI & Downsampled in-window series with day baseline & Confirm a signal's shape and anomaly \\
\addlinespace[1pt]
\texttt{get\_golden\_signal} & --- & Business-KPI anomalies and their onset & System failure timing, not location; fallback \\
\addlinespace[1pt]
\texttt{get\_call\_graph} & --- & Services ranked by latency / fail rate; call edges & Localize a network fault; trace propagation \\
\addlinespace[1pt]
\texttt{get\_log\_summary} & component & Per-log-name frequency vs.\ day baseline & Catch metric-blind faults (e.g.\ a GC burst) \\
\addlinespace[1pt]
\texttt{get\_log\_sample} & component, log name & Sampled raw log lines & Read log content to disambiguate a reason \\
\addlinespace[1pt]
\texttt{submit\_answer} & root causes & Acceptance or a correctable error & Commit and validate the final answer \\
\bottomrule
\end{tabular}
\end{table*}
\textbf{Co-location rollup.}
This step takes the cards produced so far, together with the deployment topology declared by the adapter, and adjusts the granularity at which each fault is reported.
It is needed because a fault sometimes leaves no signature on the component that actually carries it.
A node fault, for instance, may surface only as latency on the pods the node hosts, never on a node-level metric, so the raw cards point at the victims rather than the source.
EviRCA rolls evidence up in two directions.
Vertically, it groups the affected pods by their host node: when several co-located pods on one node degrade together, it emits a card for their shared node, whereas a single pod degrading while its co-located neighbors stay healthy keeps its card at the pod level.
Horizontally, it groups the affected pods by their service: when the pods of one service show the same anomaly, it emits a card for the service and removes the subsumed pod cards.
The direction and grouping are driven entirely by the declared topology, so this primitive applies to any system that declares one.
This step recovers level-spanning faults that would otherwise be reported at the wrong level, and it does so without discarding the underlying pod evidence, which stays available to the reasoning stage.

\textbf{Sub-threshold fallback.}
The detector above is deliberately high-precision, so in a few cases it surfaces no card at all.
This step guarantees that the reasoning stage is never handed an empty set: when no series crosses the detection threshold, EviRCA emits the closest sub-threshold leads, ranked by how far they deviate from their baseline.
These leads are kept separate from the high-precision cards and used only as a last resort, so they form a safety net without diluting the evidence on the cases that do produce cards.

Together, the four steps turn a window of raw, heterogeneous telemetry into a card set that is faithful, modality-complete, reported at the right granularity, and never empty, which is exactly the input the reasoning stage needs.

\subsection{Reasoning over Structured Evidence}
\label{sec:design:reasoning}

The second stage performs the one subtask that genuinely calls for an LLM: discriminating the true root cause and its reason from the competing candidates the first stage surfaced.
This is hard not because the evidence is missing but because it is correlated.
The same fault leaves traces on many components at once, so the cards alone do not say which component is the cause, which is a downstream victim, and which of several plausible reasons actually holds.
Resolving this calls for semantic judgment, which is where an LLM is uniquely useful, but EviRCA confines that judgment to a bounded, auditable interaction rather than open-ended exploration.

The stage takes the evidence cards, the query, and a small set of predefined, read-only tools, and it returns a $\langle$time, component, reason$\rangle$ tuple for each fault.
The LLM works only through these tools, and never reads raw telemetry or runs code.
It inspects the cards and their supporting signals as needed, settles on the root-cause component and reason, and submits its answer, which the system then reconciles to the number of injected faults.
The answer time is taken from the supporting card's onset rather than computed by the model, which removes a frequent source of timing error.

The tools are capability-gated: each is offered only when the system actually has the modality behind it, so the call-graph tool appears only when traces exist and the log tools only when logs do.
Through these tools, summarized in Table~\ref{tab:tools}, the model inspects the cards and their supporting signals and commits an answer.
A \emph{golden signal} here is a business-level KPI that reflects user-facing health, such as aggregate request rate, success rate, or mean response time.
It tells the model \emph{when} the system degraded, which separates a real fault from incidental noise, but not \emph{where} the fault is.

These tools let the model reason from the evidence rather than follow a fixed script, and three principles guide that reasoning.
It ranks candidates by relative magnitude rather than mere presence, because a single fault triggers anomalies in many correlated KPIs and only the relative size of a specific signature separates the cause from its effects.
It follows the call graph from the loudest victim back to the most upstream abnormal component, since the loudest signal is usually a symptom rather than the source.
And, when a golden signal exists, it uses the signal's onset to reject leads whose timing does not line up.
Because the action space is bounded and the evidence is already reduced, this stage costs far fewer tokens than an open-ended code loop and gives a stable answer across runs.

\subsection{Declarative Adapter}
\label{sec:design:adapter}

All system-specific knowledge in EviRCA resides in one declarative adapter, and the shared engine has no per-system branches.
An adapter declares only objective structure, never parameters fitted to the answers.
It names the telemetry files, their columns, and their units; the per-modality specifications that drive the detectors; the deployment topology used by the co-location rollup; the candidate hierarchy and its components; the mapping from resource type to coarse reason candidates; and an optional short paragraph of structural guidance.
Writing an adapter is a modest, one-time effort per system, and it is the only manual step needed to configure a new system.
Because both stages read their structure from the adapter, the same engine runs all three benchmark systems unchanged, despite their differences in topology, modality coverage, and telemetry encoding.
\begin{figure*}[t]
\centering
\IfFileExists{fig_case_pdf.pdf}{%
  \includegraphics[width=\textwidth]{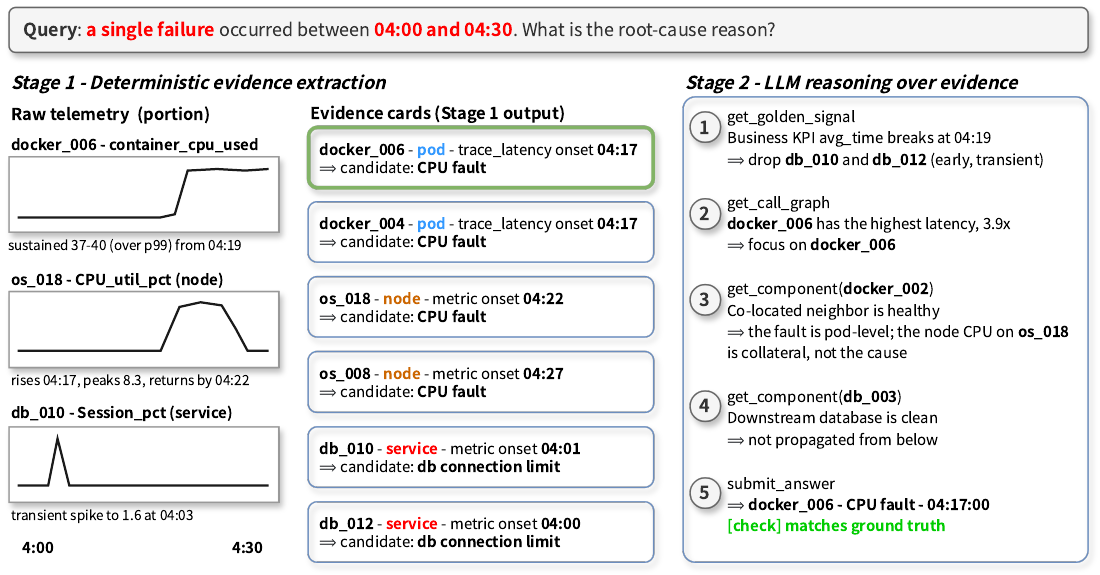}%
}{%
  \fbox{\parbox[c][3.0cm][c]{0.95\textwidth}{\centering\itshape
  [ fig\_case.pdf not found --- place the exported running-example figure beside main.tex. ]}}%
}
\caption{A running example of EviRCA on the Telecom dataset:
Stage~1 reduces raw multi-modal telemetry to evidence cards,
and Stage~2 reasons over them to pick the true root cause among the distractors.}
\label{fig:case}
\end{figure*}

\subsection{Running Example}
\label{sec:design:example}

Figure~\ref{fig:case} walks one Telecom failure through both stages.
The query asks only for the root-cause reason of a single failure between 04:00 and 04:30.

The first stage reduces the window to six evidence cards from three modalities:
two database services (\texttt{db\_010}, \texttt{db\_012}) with brief connection spikes around 04:01--04:03;
two pods (\texttt{docker\_004}, \texttt{docker\_006}) flagged by the trace modality as isolated slow pods;
and two nodes (\texttt{os\_018}, \texttt{os\_008}) with elevated CPU.
The true cause, \texttt{docker\_006}, surfaces through the trace modality, and the rollup keeps it at the pod level because its co-located neighbor \texttt{docker\_002} stays healthy.

The second stage then discriminates.
The golden signal deviates at 04:19, so the model discards the database spikes, which are transient and far too early.
The trace ranks \texttt{docker\_006} highest in latency, at a $3.9\times$ ratio.
Its container CPU stays above the 99th percentile from 04:19 onward while its co-located neighbor shows nothing.
A node-wide CPU fault would have hit both pods, so the elevated node CPU on \texttt{os\_018} is collateral, not the cause.
Inspecting \texttt{docker\_006}'s downstream database rules out propagation from below.
The model outputs \texttt{docker\_006}, CPU fault.
The first stage put the right evidence on the table, and the second stage chose correctly among the distractors.

\section{Evaluation}
\label{sec:eval}

The evaluation is framed around the following four core Research Questions (RQs):
\noindent\textbf{RQ1 (Effectiveness):} To what extent does EviRCA outperform state-of-the-art baselines in root-cause localization across structurally heterogeneous microservice systems and diverse LLM backends?
\noindent\textbf{RQ2 (Efficiency and Operational Overhead):} How does EviRCA compare to existing agentic workflows in terms of computational efficiency and wall-clock execution latency?
\noindent\textbf{RQ3 (Ablation Study):} What are the individual contributions of EviRCA's core architectural components, and to what degree are the performance gains decoupled from system-specific prompt engineering?
\noindent\textbf{RQ4 (Error Analysis):} What are the primary taxonomies and root causes of EviRCA's residual failures, and what structural bottlenecks do they reveal?

\subsection{Experimental Setup}
\label{sec:eval:setup}

\paragraph{Benchmark and systems}
OpenRCA~\cite{xu2025openrca} comprises three enterprise microservice systems that differ sharply in topology, modality coverage, and telemetry encoding.
We report them in a fixed order: \emph{Bank} (136 cases), \emph{Telecom} (51), and \emph{Market} (148).
We treat Market as one system, the union of its two deployments \texttt{cloudbed-1} and \texttt{cloudbed-2}.
Each case is a natural-language query over a 30-minute window that asks for some subset of a root cause's $\langle$time, component, reason$\rangle$.

\begin{table*}[t]
\centering
\caption{Root-cause accuracy on OpenRCA (\%).
\emph{Correct} requires all queried attributes to match; \emph{Partial} credits partially correct predictions.
\emph{Overall} is weighted by case count (Bank 136, Telecom 51, Market 148; 335 total).
Best results in \textbf{bold}.}
\label{tab:rq1-main}
{\renewcommand{\arraystretch}{1.1}
\begin{tabular}{ll cc cc cc cc}
\toprule
\multirow{2}{*}{Method} & \multirow{2}{*}{LLM}
 & \multicolumn{2}{c}{Bank} & \multicolumn{2}{c}{Telecom}
 & \multicolumn{2}{c}{Market} & \multicolumn{2}{c}{Overall} \\
\cmidrule(lr){3-4}\cmidrule(lr){5-6}\cmidrule(lr){7-8}\cmidrule(lr){9-10}
 & & Correct & Partial & Correct & Partial & Correct & Partial & Correct & Partial \\
\midrule
\multirow{2}{*}{Oracle Sampling}
 & deepseek-v4-pro & 12.5 & 20.7 & 19.6 & 25.1 & 14.9 & 22.3 & 14.6 & 22.1 \\
 & qwen3.7-plus    &  9.6 & 17.6 & 29.4 & 36.6 & 15.5 & 32.4 & 15.2 & 27.0 \\
\midrule
\addlinespace[2pt]
\multirow{2}{*}{RCA-Agent}
 & deepseek-v4-pro & 17.6 & 23.1 & 11.8 & 19.6 &  9.5 & 19.1 & 13.1 & 20.8 \\
 & qwen3.7-plus    & 22.1 & 29.1 &  9.8 & 16.0 &  6.8 & 16.3 & 13.4 & 21.5 \\
\midrule
\addlinespace[2pt]
\multirow{2}{*}{\textbf{EviRCA}}
 & deepseek-v4-pro & \textbf{52.2} & \textbf{63.2} & \textbf{62.7} & \textbf{65.4} & \textbf{29.7} & \textbf{46.7} & \textbf{43.9} & \textbf{56.2} \\
 & qwen3.7-plus    & \textbf{47.8} & \textbf{60.8} & \textbf{56.9} & \textbf{60.8} & \textbf{28.4} & \textbf{46.2} & \textbf{40.6} & \textbf{54.4} \\
\bottomrule
\end{tabular}
}
\end{table*}

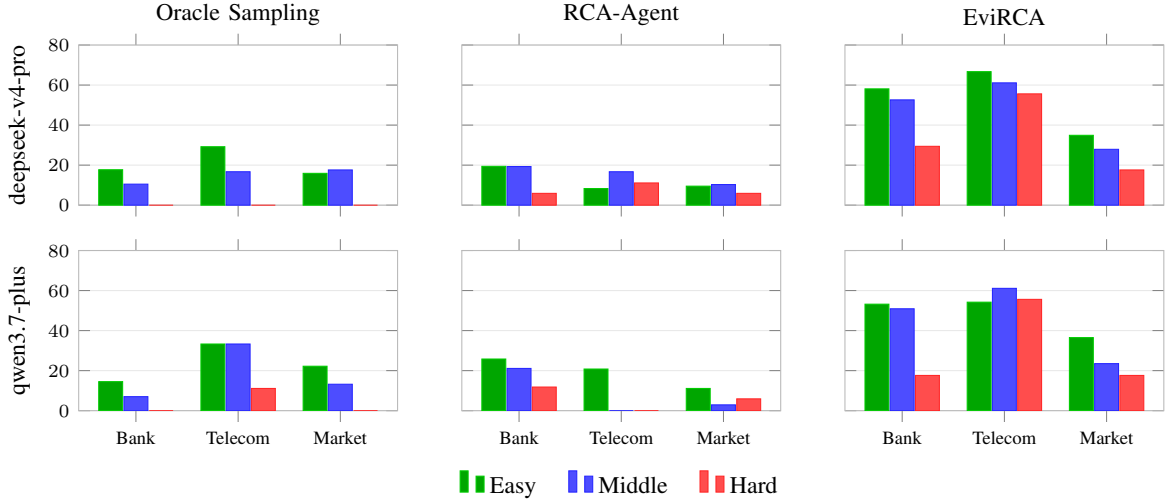
\begin{figure*}[t]
\centering
\begin{tikzpicture}
\begin{groupplot}[
  group style={
    group size=3 by 2,
    horizontal sep=0.85cm,
    vertical sep=0.6cm,
    xticklabels at=edge bottom,
    yticklabels at=edge left,
  },
  every axis/.append style={bar width=9pt},
  ybar=0.5pt,
  width=5.8cm, height=3.7cm,
  ymin=0, ymax=80,
  ytick={0,20,40,60,80},
  ymajorgrids, grid style={gray!18},
  symbolic x coords={Bank,Telecom,Market},
  xtick=data,
  enlarge x limits=0.28,
  tick label style={font=\scriptsize},
  title style={font=\small, yshift=-2pt},
  ylabel style={font=\small},
  axis line style={gray!55},
]
\nextgroupplot[title={Oracle Sampling}, ylabel={deepseek-v4-pro},
  legend to name=rqdifflegend, legend columns=3,
  legend style={draw=none, font=\small,
                /tikz/every even column/.append style={column sep=10pt}}]
\addplot[fill=green!65!black, draw=green!80!black] coordinates {(Bank,17.7) (Telecom,29.2) (Market,15.9)};
\addplot[fill=blue!70,        draw=blue!85]        coordinates {(Bank,10.5) (Telecom,16.7) (Market,17.6)};
\addplot[fill=red!70,         draw=red!85]         coordinates {(Bank,0.0)  (Telecom,0.0)  (Market,0.0)};
\addlegendentry{Easy}
\addlegendentry{Middle}
\addlegendentry{Hard}

\nextgroupplot[title={RCA-Agent}]
\addplot[fill=green!65!black, draw=green!80!black] coordinates {(Bank,19.4) (Telecom,8.3)  (Market,9.5)};
\addplot[fill=blue!70,        draw=blue!85]        coordinates {(Bank,19.3) (Telecom,16.7) (Market,10.3)};
\addplot[fill=red!70,         draw=red!85]         coordinates {(Bank,5.9)  (Telecom,11.1) (Market,5.9)};

\nextgroupplot[title={EviRCA}]
\addplot[fill=green!65!black, draw=green!80!black] coordinates {(Bank,58.1) (Telecom,66.7) (Market,34.9)};
\addplot[fill=blue!70,        draw=blue!85]        coordinates {(Bank,52.6) (Telecom,61.1) (Market,27.9)};
\addplot[fill=red!70,         draw=red!85]         coordinates {(Bank,29.4) (Telecom,55.6) (Market,17.6)};

\nextgroupplot[ylabel={qwen3.7-plus}]
\addplot[fill=green!65!black, draw=green!80!black] coordinates {(Bank,14.5) (Telecom,33.3) (Market,22.2)};
\addplot[fill=blue!70,        draw=blue!85]        coordinates {(Bank,7.0)  (Telecom,33.3) (Market,13.2)};
\addplot[fill=red!70,         draw=red!85]         coordinates {(Bank,0.0)  (Telecom,11.1) (Market,0.0)};

\nextgroupplot
\addplot[fill=green!65!black, draw=green!80!black] coordinates {(Bank,25.8) (Telecom,20.8) (Market,11.1)};
\addplot[fill=blue!70,        draw=blue!85]        coordinates {(Bank,21.1) (Telecom,0.0)  (Market,2.9)};
\addplot[fill=red!70,         draw=red!85]         coordinates {(Bank,11.8) (Telecom,0.0)  (Market,5.9)};

\nextgroupplot
\addplot[fill=green!65!black, draw=green!80!black] coordinates {(Bank,53.2) (Telecom,54.2) (Market,36.5)};
\addplot[fill=blue!70,        draw=blue!85]        coordinates {(Bank,50.9) (Telecom,61.1) (Market,23.5)};
\addplot[fill=red!70,         draw=red!85]         coordinates {(Bank,17.6) (Telecom,55.6) (Market,17.6)};

\end{groupplot}
\path (group c1r2.south west) -- (group c3r2.south east)
      node[midway, yshift=-1.0cm] {\pgfplotslegendfromname{rqdifflegend}};
\end{tikzpicture}
\caption{Strict (Correct) accuracy by system and query difficulty: three methods (columns) under two LLMs (rows). Within each panel, bars group by system and are shaded by difficulty.}
\label{fig:rq1-diff}
\end{figure*}
\paragraph{Evaluation Metric}
We use OpenRCA's scorer.
A case is \emph{Correct} only if every queried field matches the ground truth and the number of predicted root causes equals the number of injected faults.
Components and reasons must come from the system's closed candidate sets, and times must fall within $\pm 60$\,s.
This all-or-nothing \emph{Correct} rate is our primary metric.
\emph{Partial} credits the fraction of queried fields a case answers correctly.
Difficulty follows the query templates: \emph{easy} (one queried element, task 1--3), \emph{middle} (two, task 4--6), and \emph{hard} (all three, task 7).

\paragraph{Baselines}
We compare against the two methods reported for OpenRCA~\cite{xu2025openrca}, run with the same two LLMs.
\emph{RCA-Agent} is the benchmark's controller--executor agent: a planner iteratively instructs an executor that writes and runs Python over the raw telemetry in a stateful kernel.
\emph{Oracle Sampling} answers from a sampled slice of telemetry in a single LLM call. It serves as an idealized upper bound for sampling-based methods, as it draws exclusively from KPIs pre-selected by domain engineers to be diagnostic---reducing the KPI space by $95\%$ while retaining the most causally relevant signals.

\paragraph{LLM backends}
This work adopts two LLMs in different model families, including \texttt{deepseek-v4-pro} and \texttt{qwen3.7-plus}, to comprehensively test EviRCA and the comparison methods.

\subsection{RQ1: Accuracy}
\label{sec:eval:rq1}

Table \ref{tab:rq1-main} presents the experimental results among EviRCA and the baselines on OpenRCA.
It demonstrates that EviRCA outperforms both baselines across systems, LLM backends and evaluation metrics.
Specifically, EviRCA achieves a weighted overall \emph{Correct} rate of 43.9\% when paired with \texttt{deepseek-v4-pro}, and 40.6\% with \texttt{qwen3.7-plus}. 
In stark contrast, the top-performing baseline configuration yields a maximum \emph{Correct} rate of only 15.2\%, establishing a substantial performance gap where EviRCA delivers a near threefold improvement in overall accuracy.
This performance margin is particularly pronounced in the Bank and Telecom systems. 
For instance, on the Telecom dataset under identical model backends, EviRCA advances the \emph{Correct} score to 62.7\%, whereas RCA-Agent and Oracle Sampling underperform at 11.8\% and 19.6\%, respectively.
The same ordering holds on the Partial metric: EviRCA recovers 54--56\% of queried fields overall against at most 27\% for any baseline, so it gets more of the answer right even on cases it does not fully solve.

To test whether knowing \emph{where} to look is the binding constraint, we compare EviRCA against the Oracle Sampling baseline, which benefits from privileged information.
While that baseline method has prior access to a curated list of KPIs, EviRCA achieves an accuracy rate two to four times higher.
The end-to-end task comprises multiple sub-processes, and simply knowing \emph{where} to start detection is insufficient to bridge this gap.
The challenge lies not only in locating relevant telemetry data but also in identifying the fault and interpreting the root cause amidst competing evidence, which will be further discussed in Section \ref{sec:eval:rq4}.

Figure~\ref{fig:rq1-diff} breaks accuracy down by system and query difficulty.
While accuracy inherently scales down as difficulty increases from easy to hard, EviRCA consistently preserves a clear separation from the baselines at all tiers.
This contrast peaks on the hardest tier, which demands the joint prediction of the full triple, i.e., the fault's time, component, and reason together.
On these hard cases, EviRCA successfully resolves 17.6\%--29.4\% of Bank incidents and 55.6\% of Telecom incidents under both LLMs.
In comparison, prior literature reports a 0\% success rate on this hard tier across all evaluated models \cite{xu2025openrca}.
The two newer LLMs we evaluate do lift the baselines off that floor on a few cells: RCA-Agent reaches $11.8\%$ on Bank and Oracle Sampling $11.1\%$ on Telecom.
Yet no baseline cell exceeds $11.8\%$, whereas EviRCA's weakest hard cell already stands at $17.6\%$ and its strongest at $55.6\%$.
Solving the hard tier therefore calls for the joint judgment that EviRCA's design supplies, not merely a stronger model.

\subsection{RQ2: Efficiency and Operational Overhead}
\label{sec:eval:rq2}
\begin{table}[t]
\centering
\caption{Per-case cost with \texttt{deepseek-v4-pro}, where \emph{ratio} $=$ RCA-Agent\,/\,EviRCA.}
\label{tab:rq2-cost}
{\renewcommand{\arraystretch}{1.1}
\begin{tabular}{l ccc}
\toprule
 & Bank & Telecom & Market \\
\midrule
EviRCA tokens/case        & 101.8K & 48.3K & 66.4K \\
RCA-Agent tokens/case     & 1.53M  & 1.24M & 1.54M \\
\quad\emph{ratio}         & $15.0\times$ & $25.7\times$ & $23.3\times$ \\
\midrule
EviRCA time/case (s)      & 182.3 & 31.7 & 40.8 \\
RCA-Agent time/case (s)   & 577   & 622  & 685  \\
\quad\emph{ratio}         & $3.2\times$ & $19.6\times$ & $16.8\times$ \\
\bottomrule
\end{tabular}
}
\end{table}

Table~\ref{tab:rq2-cost} reports the per-case operational cost evaluated under the \texttt{deepseek-v4-pro} backend.
The empirical tracking reveals that EviRCA substantially reduces resource consumption, utilizing 15$\times$--26$\times$ fewer tokens per incident compared to RCA-Agent.
This drastic token economy stems directly from EviRCA's architectural design: it executes a single, structured reasoning pass over pre-reduced, highly concentrated evidence, thereby completely bypassing the expensive, multi-turn loops of reasoning and code execution over raw, high-volume telemetry data.

Apart from token efficiency, EviRCA achieves a 3.2× to 19.6× speedup in per-case wall-clock execution time.
We note that the latency reduction is relatively less pronounced within the Bank system.
This stems from how the Bank system encodes traces: they carry no call status or explicit callee and must be reassembled from per-span parent links, over the largest trace volume of the three systems, so EviRCA reconstructs the relevant call graph separately for each case.
Telecom and Market do not incur this overhead, since their traces can be summarized once upfront, into a compact per-minute call summary that every case reuses, leaving per-case trace work negligible.

Overall, EviRCA's efficiency advantage holds across all three systems.
The token reduction, the dominant and model-independent cost driver, remains at an order of magnitude or more ($15$--$26\times$) on every system, while wall-clock time drops by $3.2$--$19.6\times$.
Even on Bank, where per-case trace reconstruction narrows the time margin, EviRCA remains both substantially cheaper and faster than the agentic baseline.

\subsection{RQ3: Ablation Study}
\label{sec:eval:rq3}

\begin{table}[t]
    \centering
    \caption{The impact on the \emph{Correct} rate (in percentage points) via leave-one-out module removal, which is evaluated with \texttt{deepseek-v4-pro}.}
    \label{tab:rq3}
    {\renewcommand{\arraystretch}{1.1}
    \begin{tabular}{l rrr}
        \toprule
        Module removed      & Bank   & Telecom & Market \\
        \midrule
        Trace               & $-3.7$ & $-45.1$ & $-0.7$ \\
        Co-location rollup  & --     & $-37.3$ & $-1.4$ \\
        Per-system guidance & $+0.7$ & $-11.8$ & $0.0$  \\
        Golden signal       & $0.0$  & $-3.9$  & $0.0$  \\
        Safety net          & $-2.9$ & $0.0$   & $-3.4$ \\
        Log modality        & $-1.5$ & --      & --     \\
        \bottomrule
    \end{tabular}
    }
\end{table}

To investigate the individual contributions of EviRCA's core components, we conduct a leave-one-out ablation study using the \texttt{deepseek-v4-pro} backend.
Table~\ref{tab:rq3} reports the deviation in the Correct rate when each module is disabled. Crucially, the performance impact of each component is highly system-specific, closely aligning with the underlying telemetry characteristics and fault distributions of each target system.

On the Telecom system, disabling either the trace modality or the co-location rollup leads to a catastrophic performance drop, i.e., $-45.1$ and $-37.3$ percentage points in the Correct rate, respectively.
These two primitives enable EviRCA to localize Telecom's node faults, which leave little or no signature on the node's metric: such a fault surfaces as latency on the node's co-located pods through the traces, and the co-location rollup is what aggregates those slow pods back to the responsible node.
Without either, the node fault cannot be reported at its correct level.
On Bank, by contrast, no single module dominates: the trace modality, the safety net, and the log modality each move the Correct rate by only a few points ($-3.7$, $-2.9$, and $-1.5$).
This reflects a system whose fault signatures are spread across resource types and modalities rather than concentrated in one primitive.

The per-system guidance consists of a short hand-written paragraph, and it is the only point where expert knowledge about a system enters the pipeline.
We call attention to it because, if EviRCA’s advantage stemmed from prompt tuning to the benchmark, removing this guidance should cause accuracy to collapse.
It does not: removing the guidance impairs performance only on Telecom (-11.8 points, where propagation reasoning matters most); on Market it has no effect (0.0), and on Bank it even yields a slight improvement (+0.7 points).
These gains are therefore architectural, not an artifact of prompt tuning.
Finally, as discussed in Section~\ref{sec:eval:rq4}, Market’s bottleneck lies upstream of its modules; accordingly, no single ablation alters its Correct rate by more than three percentage points.

\subsection{RQ4: Failure Analysis}
\label{sec:eval:rq4}

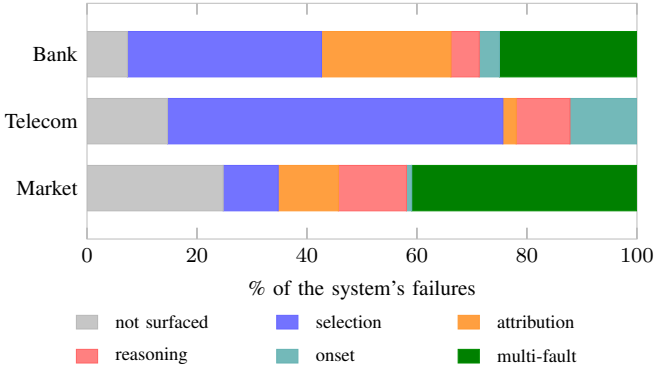
\begin{figure}[t]
  \centering
  \begin{tikzpicture}
    \begin{axis}[
        width=\columnwidth, height=4.7cm,
        xbar stacked,
        xmin=0, xmax=100,
        xlabel={\footnotesize \% of the system's failures},
        symbolic y coords={Market,Telecom,Bank},
        ytick=data,
        bar width=6mm,
        enlarge y limits=0.38,
        tick label style={font=\footnotesize},
        label style={font=\footnotesize},
        axis line style={gray!50},
      ]
      \addplot[fill=gray!45,draw=gray!55]              coordinates {(7.35,Bank)(14.63,Telecom)(24.76,Market)};
      \addplot[fill=blue!55,draw=blue!60]              coordinates {(35.29,Bank)(60.98,Telecom)(10.00,Market)};
      \addplot[fill=orange!75,draw=orange!80]          coordinates {(23.53,Bank)(2.44,Telecom)(10.95,Market)};
      \addplot[fill=red!50,draw=red!60]                coordinates {(5.15,Bank)(9.76,Telecom)(12.38,Market)};
      \addplot[fill=teal!55,draw=teal!60]              coordinates {(3.68,Bank)(12.20,Telecom)(0.95,Market)};
      \addplot[fill=green!50!black,draw=green!55!black] coordinates {(25.00,Bank)(0.00,Telecom)(40.95,Market)};

    \end{axis}

    \begin{scope}[shift={(0,-0.55)}]
      \node[fill=gray!45,  draw=gray!55,  minimum width=0.28cm, minimum height=0.18cm, inner sep=0pt] (b1) at (0.00cm, -0.55cm) {};
      \node[anchor=west, font=\scriptsize] at ([xshift=3pt]b1.east) {not surfaced};

      \node[fill=blue!55,  draw=blue!60,  minimum width=0.28cm, minimum height=0.18cm, inner sep=0pt] (b2) at (2.65cm, -0.55cm) {};
      \node[anchor=west, font=\scriptsize] at ([xshift=3pt]b2.east) {selection};

      \node[fill=orange!75,draw=orange!80,minimum width=0.28cm, minimum height=0.18cm, inner sep=0pt] (b3) at (5.05cm, -0.55cm) {};
      \node[anchor=west, font=\scriptsize] at ([xshift=3pt]b3.east) {attribution};

      \node[fill=red!50,   draw=red!60,   minimum width=0.28cm, minimum height=0.18cm, inner sep=0pt] (b4) at (0.00cm, -1.00cm) {};
      \node[anchor=west, font=\scriptsize] at ([xshift=3pt]b4.east) {reasoning};

      \node[fill=teal!55,  draw=teal!60,  minimum width=0.28cm, minimum height=0.18cm, inner sep=0pt] (b5) at (2.65cm, -1.00cm) {};
      \node[anchor=west, font=\scriptsize] at ([xshift=3pt]b5.east) {onset};

      \node[fill=green!50!black,draw=green!55!black,minimum width=0.28cm, minimum height=0.18cm, inner sep=0pt] (b6) at (5.05cm, -1.00cm) {};
      \node[anchor=west, font=\scriptsize] at ([xshift=3pt]b6.east) {multi-fault};
    \end{scope}
  \end{tikzpicture}
  \caption{Failure profile by the pipeline stage at which each failure occurs, \texttt{deepseek-v4-pro} and \texttt{qwen3.7-plus}.
    \emph{not surfaced} = the true component never reached the evidence (a recall gap);
    \emph{selection}/\emph{attribution}/\emph{reasoning} are wrong decisions over \emph{already-surfaced} evidence;
    \emph{multi-fault} cases are broken out separately.}
  \label{fig:rq4-mech}
\end{figure}
\begin{table}[t]
\centering
\caption{
Distribution and overlap of failure cases between \texttt{deepseek-v4-pro} and \texttt{qwen3.7-plus}, where $\mathit{shared}$ represents the percentage of cases failed by both models out of all cases failed by at least one model.
}
\label{tab:rq4-overlap}
{\renewcommand{\arraystretch}{1.2}
\begin{tabular}{l rrr r}
\toprule
System & both-fail & one model only & both-pass & shared \\
\midrule
Bank    & 61 & 14 & 61 & 81.3\% \\
Telecom & 19 &  3 & 29 & 86.4\% \\
Market  & 97 & 16 & 35 & 85.8\% \\
\bottomrule
\end{tabular}
}
\end{table}

We categorize each failure case by both its specific breakdown stage within the pipeline (Fig.~\ref{fig:rq4-mech}) and its cross-model consistency (Table~\ref{tab:rq4-overlap}). 
This fine-grained categorization yields three key insights that refine our core empirical findings.

\textbf{Most residual failures stem from erroneous decisions over surfaced evidence rather than a lack of coverage.}
Among single-fault failures, the extraction stage fails to surface the true component in only about a quarter of cases (input-missing $22.5\%$).
The rest occur on evidence that \emph{was} surfaced: a downstream component with a stronger anomaly chosen over the true root cause ($35\%$), the wrong reason among the candidates ($21\%$), or an incorrect propagation chain ($14\%$).
This matches the privileged Oracle Sampling result from RQ1: being given where to look does not settle which candidate is the cause.

\textbf{Failures are largely shared across models, which explains the narrow spread between LLMs.}
$81$--$87\%$ of failing cases fail under \emph{both} models (Table~\ref{tab:rq4-overlap}), and only $13$--$19\%$ are model-specific.
The residual error is dominated by genuinely ambiguous decisions, such as network delay versus packet loss, or a propagated symptom versus its source, that swapping models within this capability tier does not resolve.
This explains why the two LLMs perform so similarly in Table~\ref{tab:rq1-main}: within the same capability tier, a stronger model cannot resolve the ambiguity that the evidence inherently leaves.

\textbf{The systems have distinct failure profiles} (Fig.~\ref{fig:rq4-mech}).
On Telecom, failures are almost entirely selection errors ($61\%$): the correct node is usually surfaced in the evidence as a low-confidence lead, but a competing lead is chosen instead.
Bank splits across selection, attribution, and multi-fault.
Market is dominated by \emph{multi-fault} ($41\%$) and by \emph{un-surfaced} evidence ($25\%$), and that un-surfaced evidence is concentrated in network faults.
Among Market's failing network-reason cases, the true component never reaches the evidence in $89.6\%$ of them, versus $8.5\%$ on Bank and $0\%$ on Telecom.
This is a coverage limit of the evidence layer for network signals rather than a reasoning failure, and we return to it among the limitations.

\subsection{The Per-System Picture}
\label{sec:eval:summary}

Reading the four questions by system, rather than by question, clarifies the overall picture.
On Bank, EviRCA is accurate ($52.2\%$) but its modules are individually small (RQ3), and its failures split across selection, attribution, and multi-fault (RQ4), so the bottleneck is judgment rather than coverage.
On Telecom, EviRCA is strongest ($62.7\%$ Correct) and most clearly model-driven: trace and the co-location rollup are decisive (RQ3), and the residual error is almost entirely selection among surfaced leads (RQ4).
On Market, EviRCA is weakest ($29.7\%$), and its modules are nearly inert (RQ3) because the binding limits lie upstream: a large multi-fault share, and network faults that the evidence layer does not surface (RQ4).
The same engine therefore succeeds or struggles for system-specific reasons, and those reasons are coverage and ambiguity rather than a failure of the reasoner.

\section{Discussion and Limitations}
\label{sec:discussion}

\subsection{Further Discussions} 
The results show that, for microservice RCA, the key issue is to draw a proper boundary between deterministic evidence extraction and LLM reasoning. EviRCA assigns anomaly detection and candidate localization to a deterministic stage. It then lets the LLM infer the root cause from structured evidence cards. This constrained design improves the overall Correct rate to 40.6\%-43.9\%, while the best baseline reaches only 15.2\%. It also uses 15-26 times fewer tokens and reduces per-case latency by 3.2-19.6 times. These results suggest that free-form exploration over raw telemetry is not necessary in this setting. It can also introduce context expansion, schema hallucination, and run-to-run instability.
The failure analysis further shows that most residual errors come from incorrect judgments over extracted evidence. Among single-fault failures, the true component is missing from the extracted evidence in only 22.5\% of cases. For example, the model may select a downstream victim with a stronger anomaly as the root cause, choose the wrong reason among similar candidates, or misinterpret the propagation chain. In addition, 81\%-87\% of failures are shared by the two LLMs. This suggests that the remaining difficulty is not easily solved by replacing the model within the same capability tier. Instead, the main challenge is to make the correct interpretation among correlated and interfering evidence.

\subsection{Limitations}
The main limitation of EviRCA is that the evidence layer determines its performance ceiling. If a true fault signal is not included in the evidence cards due to missing modalities, coarse granularity, or unsuitable detection rules, the downstream LLM cannot recover it. In failing network-reason cases, the true component is absent from the evidence in 89.6\% of Market cases, compared with 8.5\% in Bank and 0\% in Telecom. This helps explain why Market has lower accuracy than the other two systems. Future work may address this issue by adding richer modalities, using finer-grained features.
In addition, EviRCA inherits several assumptions from OpenRCA, such as closed sets of candidate components and reasons, known fault counts, and available topology information. These assumptions may not always hold in real production environments.

\section{Related Work}
\label{sec:related}

\subsection{Root Cause Analysis for Microservices}
As discussed previously, root-cause analysis over microservice telemetry is commonly decomposed into anomaly detection, localization of faulty components, and attribution of failure time, affected component, and underlying reason.
Prior to recent LLM-based work, microservice RCA primarily relied on localization methods grounded in distributed traces, metric time series, and service dependency graphs.
Representative approaches include random-walk ranking over service call graphs~\cite{10.1145/2465529.2465753}, causal-graph construction~\cite{10.1007/978-3-030-03596-9_1, 9110353}, trace analysis~\cite{9521340}, causal inference and discovery~\cite{10.1145/3534678.3539041, 3600270.3602529, zhang2025dynacausaldynamiccausalityawareroot}, deep-learning-based QoS debugging~\cite{10.1145/3297858.3304004}, and multimodal localization~\cite{10.1145/3611643.3616249, 10172617}.
Although effective, these methods typically require labeled service topology, supervised training data, or deployment-specific model calibration, limiting their transferability across systems.

\subsection{LLM-based RCA Agents}
Recent work has applied LLMs to RCA along two lines.
One line operates on textual incident artifacts rather than raw telemetry, producing incident summaries, coarse root-cause categories, or mitigation recommendations from historical reports~\cite{10172904, 10.1145/3627703.3629553, 10.1145/3597503.3639081}; these outputs support triage but do not localize faults within live telemetry.
Another line gives the model direct control over telemetry analysis.
Tool-augmented and code-generating agents retrieve and analyze telemetry autonomously~\cite{10.1145/3627673.3680016, 10.1145/3663529.3663841, qiao2024taskweavercodefirstagentframework}, with later systems adding SOP-guided multi-agent orchestration~\cite{10.1145/3701716.3715225}, graph-augmented workflows~\cite{tian2025galagraphaugmentedlargelanguage}, multimodal tool invocation~\cite{11229957}, and meta-level causal knowledge~\cite{liang2026metarcageneralizablerootcause}.
RCA-Agent is representative of this direction, allowing an LLM to iteratively write and execute Python over raw telemetry in a ReAct loop~\cite{xu2025openrca, yao2023react}.
These approaches therefore place substantial autonomy in the model: the LLM chooses what evidence to inspect and often generates code to inspect it.
A smaller line of work pre-structures telemetry before model invocation~\cite{luo2026opsagentevolvingmultiagentincident, Di2025OpenDeriskAI}, but still processes the resulting evidence through autonomous multi-agent loops.

\subsection{LLM Agents for Software Engineering}
Beyond RCA, LLM agents have advanced code generation, issue resolution, and test generation through benchmarks such as SWE-bench~\cite{jimenez2024swebench} and systems such as SWE-agent~\cite{3737916.3739517}, OpenHands~\cite{wang2025openhands}, and MetaGPT~\cite{hong2024metagpt}.
This progress relies on natural-language-friendly inputs and inexpensive test-based verification, which turn open-ended exploration into a checkable loop.
Even in that setting, constrained action spaces often outperform broader autonomy, as shown by structured code search~\cite{zhang2024autocoderover}, localize-then-repair pipelines without autonomous control flow~\cite{xia2025demystifying}, and constrained agent interfaces~\cite{3737916.3739517}.
Telemetry-grounded RCA lacks both conditions: its inputs are heterogeneous numeric signals, and predicted root causes have no cheap executable check.
Code generation over raw telemetry therefore inherits the cost of open-ended exploration without the test oracle that makes such exploration effective in software-engineering tasks.

\section{Conclusion}
\label{sec:conclusion}

We presented EviRCA, a framework that addresses the mismatch between LLMs and raw microservice telemetry by decoupling deterministic evidence extraction from LLM-based reasoning.
A system-agnostic extraction stage reduces heterogeneous metrics, traces, and logs into compact, faithful multimodal evidence cards, with system-specific structure isolated in a lightweight declarative adapter.
The LLM then diagnoses the root cause only through a restricted, read-only interface over these cards, without accessing raw telemetry or executing code.
Across three enterprise systems and two LLMs, EviRCA achieves a Correct rate of 40.6--43.9\%, compared with at most 15.2\% for the strongest OpenRCA baseline under matched models.
It also reduces token consumption by 15--26$\times$ and per-case latency by 3.2--19.6$\times$, while solving hard-tier cases requiring the complete time, component, and reason tuple.
Through ablation and process-level failure analysis, we show that the main bottleneck is not open-ended search over raw telemetry, but judgment over evidence that has already been surfaced.
At the same time, cases such as Market network faults show that the evidence layer still determines the performance ceiling when key signals are missing.
These findings make the extraction--reasoning boundary the central design lever for LLM-based RCA, and point to richer, finer-grained, and adaptive evidence extraction as the main route to further progress.

\section{Data Availability}
\label{sec:data-availability}

To support reproducibility, our implementation, experiment scripts, and a README with reproduction instructions are publicly available at
\url{https://github.com/yuhao541/EviRCA}.
Experiments are performed with the publicly available OpenRCA benchmark~\cite{xu2025openrca}; instructions for obtaining it
are included in the repository.

\bibliographystyle{IEEEtran}
\bibliography{main}

\end{document}